\documentclass[a4paper,12pt,oneside]{article}
\usepackage[english]{babel}
\usepackage{graphicx}

\begin{document}

\title {On the origin of continuum and line emission in CTTSs}

\author{S.A. Lamzin$^1$, M.M. Romanova$^2,$ A.S. Kravtsova$^1$}

\date{
$^1$ Sternberg Astronomical Institute,
Universitetskij prospect 13, Moscow, 119991, Russia;
{\it lamzin@sai.msu.ru, kravts@sai.msu.ru}
\\
$^2$ Department of Astronomy, Cornell University, Ithaca, NY
14853-6801, USA; {\it romanova@astro.cornell.edu} \\
}

\maketitle

\begin{abstract}
  We calculated profiles of CIV\,1550, Si\,IV 1400, NV\,1240 and OVI\,1035
doublet lines using results of 3D MHD simulations of disc accretion onto
young stars with dipole magnetic field. It appeared that our
calculations can not reproduce profiles of these lines observed 
(HST/GHRS-STIS and FUSE) in CTTSs's spectra. We also found that the 
theory predicts much larger C\,IV 1550 line flux than observed (up to 
two orders of magnitude in some cases) and argue that the main 
portion of accretion energy in CTTSs is liberated outside accretion 
shock. We conclude that the reason of disagreement between the 
theory and observation is strongly non-dipole character of CTTS's
magnetic field near its surface.
\footnote{
Published in Proc. of IAU Symp. 243: "Star-Disk Interaction in Young Stars",
Grenoble, France, 2007, Eds. J. Bouvier \& I. Appenzeller, p.115
}
\end{abstract}

\section{Introduction}

  Since the beginning of the 1990s, there has been a consensus that the line
and continuum emission observed in the spectra of classical T Tauri stars
(CTTSs) results from the magnetospheric accretion of circumstellar material.
More precisely, the magnetic field of the star is believed to stop the
accretion disk from reaching the stellar surface. In some way the
disk material becomes frozen in the magnetospheric field lines and slides 
along them toward the stellar surface, eventually being accelerated to
velocities $V_0 \sim 300$ km/s. The gas is then decelerated in an accretion
shock (AS), whose radiation presumbly gives rise to the observed line and
continuum emission.

\begin{figure}
 \begin{center}
  \includegraphics[height=7cm]{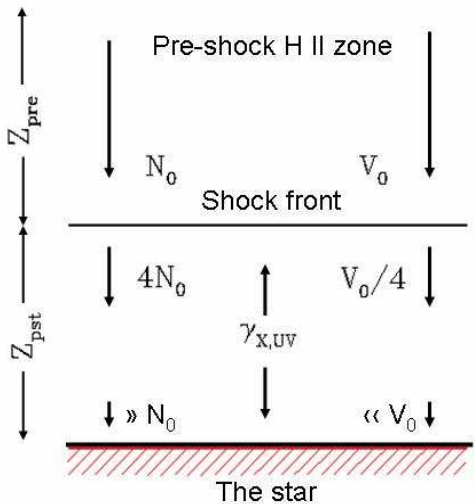}
  \caption{Schematic structure of CTTS's AS.}
  \label{fig:as-scheme}
 \end{center}
\end{figure}

  Radial extention of pre- and post-shock radiating regions of CTTS's AS
is much smaller than stellar radius -- $Z_{pre}, Z_{pst} \ll R_*$ in 
figure~\ref{fig:as-scheme}, -- making it possible to calculate the 
structure and spectrum of the AS in 1-D approximation (Lamzin, 1995).
Calculations of Lamzin (1998) and Calvet \& Gullbring (1998)
indicated that the structure of the flow can be specified nearly
unambiguously by two parameters: the velocity $V_0$ and density $\rho_0$ 
(or particle number density $N_0)$ of the gas far in front of the shock.

  Calvet \& Gullbring (1998) used results of their calculations to derive the
parameters of the accretion shock via modeling of the continuum spectral
energy distributions of classical T Tauri stars. However, they did not took
into account limb darkening effect and what is more the agreement between
the calculated and observed spectra of the veiling continuum cannot be
considered as a decisive support for the magnetospheric model, since 
boundary-layer models provide equally good agreement -- see e.g.
Basri \& Bertout (1989). The line spectrum is far more informative, and
comparisons of the calculated and observed intensities and profiles of
emission lines enable detailed studies of the accretion processes.

  Optically thin lines are best suited for this purpose: their intensity
ratios can be used to derive physical conditions independent of the geometry
of the region where they are formed, and the line profiles provide
information about both the velocity field and geometry of this region. The
calculations of Lamzin \& Gomez de Castro (1999) demonstrated that the
O\,III] 1663, Si\,III] 1892 and C\,III] 1909 lines should display the
highest intensities among the optically thin lines, and just these lines
were used to determine the accretion-shock parameters for several young
stars in that paper. However, Gomez de Castro \& Verdugo (2001, 2007)
questioned whether these lines form in the AS. Spectral lines of neutral or
singly ionized atoms apparantly form not only in AS but in magnetospheric
flow and wind as well -- see e.g. Edwards (2007) -- and thus likewise
cannot be used for diagnostic of AS.

  Resonant UV lines of the C\,IV, Si\,IV, N\,V and O\,VI uv1 doublets looks
as the most suitable lines for diagnostic of CTTSs's AS, especially lines 
of C\,IV 1550 doublet: they are strong in CTTS's spectra and calculation 
of their intensities is relatively simple.

\section{Intensities of C\,IV 1550 doublet lines:
theory vs. observations}\label{sec:c4relint}

  Generally speaking, lines of C$^{+3},$ Si$^{+3},$ N$^{+4}$ and O$^{+5}$
ions form before and behind the shock front. Gas temperature in the
pre-shock (precursor) zone of CTTSs does not exceed 20.000 K, but ions
up to O$^{+5}$ (at $200 <V_0<400$ km/s) exist here due to photoionization of
accreted matter by X-ray and UV quanta from post-shock cooling zone. When
infalling gas crosses the shock front its temperature raises up to 1-3 MK 
and for example C$^{+3}$ ions almost complitely transform to C$^{+6}$ ions.
Then gas cools and ions of interest appeares again but at that moment
gas velocity is very close to zero -- see Lamzin (1998) for details.

\begin{figure}
 \begin{center}
  \includegraphics[height=7cm]{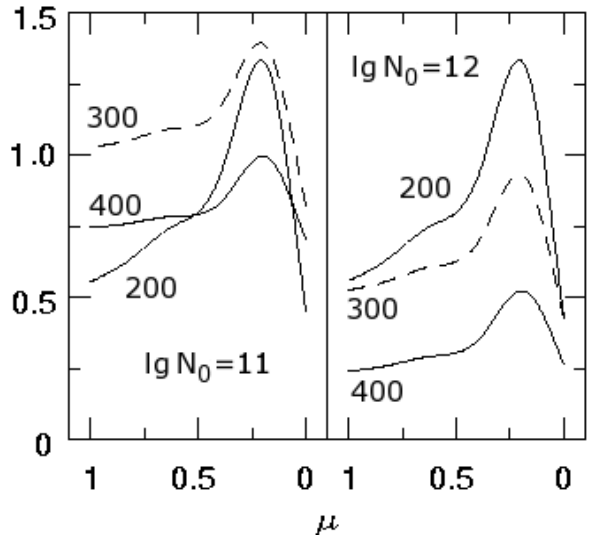}
  \caption{Relative intensities of C\,IV 1550 doublet lines
expressed in \%. See text for details.
}
  \label{fig:delta-c4}
 \end{center}
\end{figure}

  Lamzin (2003a) carried out non-LTE calculations of profiles of C\,IV 1550, Si\,IV 1400,
N\,V 1240 and O\,VI 1035 doublet lines for a plane-parallel shock viewed at
various angles. Calculations were performed for the range of preshock gas
parameters $V_0,$ $N_0$ appropriate for CTTSs. Intensities of C\,IV
1548+1551 lines, normalized to ${\cal F}=\rho_0 V_0^4/4$ value, as a function
of cosine $\mu$ of an angle between the normal to the shock front and the
line of site are presented in figure~\ref{fig:delta-c4}. The ratio is
expressed in percent, such as different lines in fugure respect to 
different infall gas velocities $(V_0=200,$ 300 and 400 km/s). Presented
results were calculated for gas particle density $N_0=10^{11}$ cm$^{-3}$ 
(left panel) and $10^{12}$ cm$^{-3}$ (right panel).

  The value ${\cal F}$ was chosen for normalisation because just this
value expected to be equal to the bolometric flux of veiling continuum
emission produced by AS. Indeed, according to current paradigm a half
of X-ray and UV quanta from post-shock cooling zone moves to the stellar
surface. These quanta are absorbed in upper layers of stellar atmosphere
and then should be reradiated predominantly in the continuum -- see
Calvet \& Gullbring (1998) for details. Thus the ratio $\delta,$ depicted
in figure~\ref{fig:delta-c4}, is theoretical prediction for ratio of C\,IV 
1550 doublet line flux to bolometric flux of veiling continuum. Thus
this ratio expected to be $\simeq 1\,\%$ almost independenly on
$V_0, N_0$ parameters of AS.

  Meanwile it was found that observed ratio is much smaller -- see
table~\ref{tab:AS-delta} in which we summarised results of our analyses of
CTTSs UV spectra observed from Hubble Space Telescope (Kravtsova \& Lamzin,
2002a,b; Kravtsova, 2003; Lamzin et~al., 2004). The descripancy between the
theory and observation is significant -- more than two orders of magnitude
in the case of RY Tau and DR Tau, such as this conclusion does not depend on
current uncertainty of value and law of interstellar extinction in the
direction to investigated stars.

\begin{table}[h!]\def~{\hphantom{0}}
  \begin{center}
  \caption{Relative contribution of C\,IV lines in emission of AS}
  \label{tab:AS-delta}
  \begin{tabular}{l|cccccc|c}\hline
Star: & RY Tau & DR Tau & T Tau & DS Tau & BP Tau & DG Tau & Theory \\\hline
$\delta,\,\%$  & 0.002  & 0.003  & 0.02~ & 0.02~  & 0.04~  & 0.07~  & 
$\sim 1.0$~\\\hline
  \end{tabular}
 \end{center}
\end{table}

  We suppose that this discrepancy means that the main portion of
veiling continuum (up to 99 \% in some cases juging from 
table~\ref{tab:AS-delta}) originates outside C\,IV 1550 line formation 
region, i.e. outside {\it strong} (Mach number $M_{sh} \gg 1)$ AS. 
In other words we conclude that the main portion of accreted matter 
does not pass throught the AS and falls to the star almost parallel 
to the stellar surface. In this part of accretion flow transformation 
of kinetic energy of infalling gas into the heat and then into radiation 
should occur in the same way(s) as in a boundary layer, i.e in a series 
of {\it weak} ($M_{sh} \simeq 1)$ oblique shocks. One can 
estimate the maximal possible angle $\gamma_{sh}$ between front of
the weak oblique shocks and stellar surface as follows.

  In the coronal equlibrium approximation C$^{+3}$ ions forms at $T\simeq
10^5$ K. Such temperature can be reached immediately behind the shock front
if accreted gas velocity component $V_r$ normal to stellar surface is
$\simeq 70$ km/s. Therefore oblique ASs with $V_r$ less than this value 
can not contribute to C\,IV 1550 line emission of accretion flow
but produce veiling continuum and emission in lines of neutral, singly
or twice ionized atoms. If typical infall gas velocity $V_0$ is $\simeq 300$
km/s, then $\gamma_{sh} = \sin^{-1}(V_r/V_0) < 15^o,$ i.e. gas producing 
such weak shocks indeed falls to the star almost parallel to its surface.

\section{Profiles of C\,IV 1550 doublet lines:
theory vs. observations}\label{sec:c4prof}

 As was mentioned above CIV\,1550 doublet lines form in two spatially 
distinct regions of strong AS: in the radiative precursor and in
the post-shock zone. Gas velocity in these regions are different: $V\simeq 
V_0 \sim 300$ km/s in the pre-shock zone
and $\sim 5-10$ km/s in the post-shock line formation region. As a result
profile of e.g. CIV\,1548 line in the spectrum of plane-parallel shock,
viewed from the direction perpendicular to the surface of shock front,
should have two components: the first one is almost at zero-velocity position
and the second is redshifted to $V_0.$ As follows from our calculations
(Lamzin 2003a), both components are optically thick, resulting in
FWHM of each componet $\sim 20-30$ km/s, with relative strength depending
on $V_0:$ at $V_0<300$ km/s "zero-velocity" component is stronger than
"high-velocity" one and vice versa at $V_0>300$ km/s. If the shock is 
viewed from the direction, that makes an angle $\theta$ with the 
perpendicular to the shock's surface, then "zero-velocity" 
component should be seen practically at the same position but the redshift
of "high-velocity" peak should be now $V_0\,\mu,$ where $\mu=\cos \theta.$ 
The same is true (in a qualitative way) for lines of Si\,IV 1400, NV\,1240
and OVI\,1035 doublets.

\begin{figure}[hb!]
 \begin{center}
  \includegraphics[height=8cm]{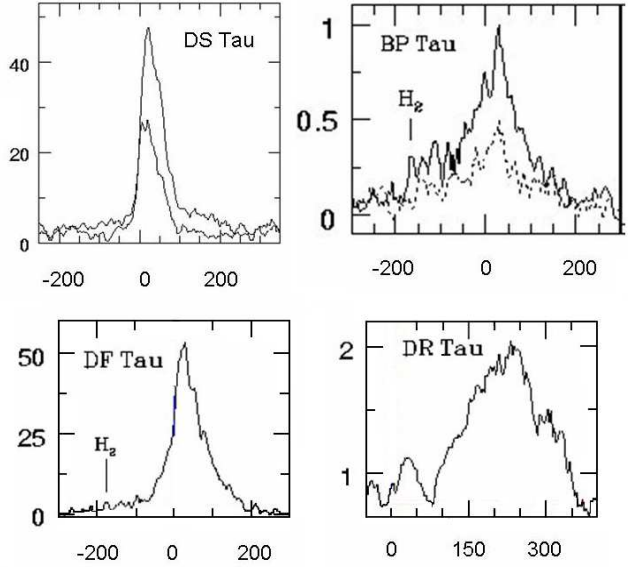}
  \caption{Profiles of C\,IV 1550 doublet lines in spectra of some CTTSs.}
  \label{fig:c4-obsprof-1}
 \end{center}
\end{figure}

  Consider now a part of CTTS's surface occupied with strong AS
(accretion zone). Observed profile of e.g. CIV\,1548 line emitted by AS
is a sum (an integral) of double-peaked profiles from all elementary area 
$\Delta S$ of the accretion zone (multiplied to $\mu\,\Delta S$ factor).
All elementary areas are viewed at different angles due to curvature of 
stellar surface and these angles varies with time due to stellar rotation.
One can expect that intensities of "zero-velocity" components from all 
parts of accretion zone will be summarised and the (weighted) 
sum of high-velocity components will results in more or less wide red 
wing or separated redshifted component depending on distribution of $V_0,$ 
$N_0$ parameters in accretion zone and on its geometry. Obviously the 
profile should vary with time due to stellar rotation and non-stationary 
accretion as well.

 Lamzin (2003b) calculated profiles of C\,IV 1550 doublet lines 
from strong AS assuming that: 1) matter falls to the star in radial direction;
2) $V_0$ and $N_0$ parameters of the shock are constant within accretion zone; 
3) the zone has the shape of circular spot or sperical belt. Results of
the calculations were compared with profiles of the lines in
UV spectra of CTTSs observed with Goddard High Resolution Spectrograph (GHRS)
and Space Telescope Imaging Spectrograph (STIS). Observational data
were extracted from Scientific Archive of Hubble Space Telescope
(http://archive.stsci.edu/hst/\\target\_descriptions.html). 
Calculated profiles differs significantly from observed ones presumbly 
because our assumptions about the character of accretion flow 
near stellar surface were not realistic enough.

\begin{figure}[ht!]
 \begin{center}
  \includegraphics[height=8cm]{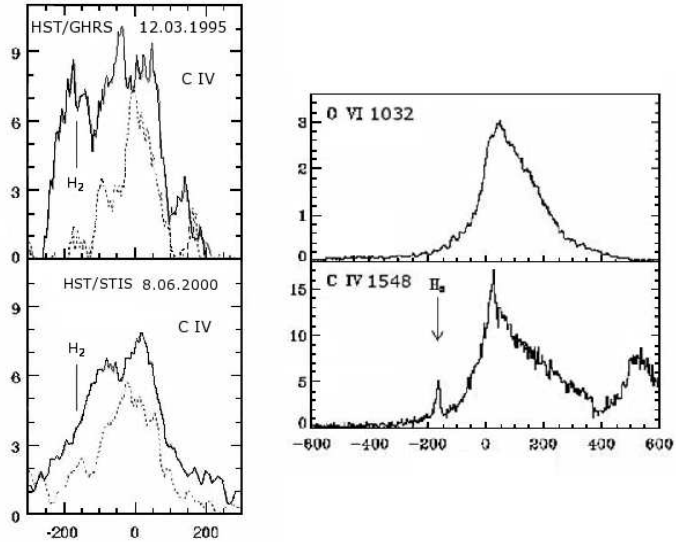}
  \caption{Profiles of C\,IV 1550 and O\,VI 1035 doublet's components
in spectra of T Tau (left panel) and TW Hya (right panel).}
  \label{fig:c4-obsprof-2}
 \end{center}
\end{figure}

  One can expect better agreement if to use parameters of accretion flow
derived from 3D MHD simulations of disc accretion to a slowly rotating 
magnetized young star with its dipole moment inclined at an angle $\alpha$ 
to the stellar rotation axis. Accretion rate $\dot M_{ac},$ polar
magnetic field strength $B$ as well as mass and radius of the central star
are free parameters of these simulations in addition to the angle 
$0 \le \alpha \le 90^o$ -- see Romanova et al. (2003) for details.
Velocity field ${\bf V}_0$ and gas density $\rho_0$ at stellar surface,
adopted from the simulations, we used as input parameters to calculate 
profiles of CIV\,1550, Si\,IV 1400, NV\,1240 and OVI\,1035 doublet lines. 
For all models we adopted $M_*= 0.8$ M$_\odot,$ $R_* = 1.8$ R$_\odot,$ 
$B=1-3$ kG and varied $\alpha,$ $\dot M_{ac}$ parameters in the 
$0^o-90^o$ and $10^{-8}-3\cdot 10^{-7}$ M$_\odot$/yr intervals respectively.
Profiles were calculated for each accretion zone's model with different 
values of an angle between stellar rotation axis and the line of 
sight $(0^o \le i \le 90^o)$ as well as for a set of phases of stellar 
rotation periods, i.e. for different angles $\psi$ (in $2\pi$ units)
between magnetic dipole axis and the plane, which contains rotation axis 
and the line of sight $(0\le \psi \le 1).$

  Observed profiles of C\,IV 1550 doublet lines in DS Tau, BP Tau, DF Tau
and DR Tau spectra are shown in figure~\ref{fig:c4-obsprof-1}. Solid and
dashed lines depicts C\,IV 1548 and C\,IV 1551 componens of the doublet.
Profiles of C\,IV 1550 doublet's components in spectra of T Tau are shown in
figure~\ref{fig:c4-obsprof-2} (left column). T Tau is the only star where
there is more than one high resolution UV spectrum and one can observe
variability of C\,IV doublet lines profiles. Only in the case of TW Hya
(right panel of the figure) there is possibility to obtain information about
lines of O\,VI 1035 doublet -- see Ardila (2007) for reference and
details.

\begin{figure}[h!]
 \begin{center}
  \includegraphics[height=12cm]{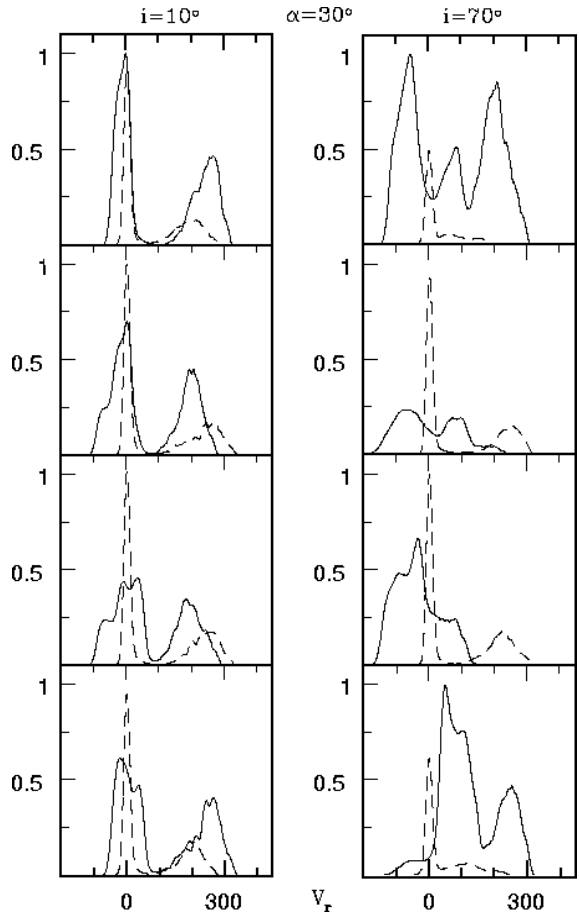}
  \caption{Theoretical profiles of C\,IV\,1548 line calculated
for CTTS with dipole magnetic field, axis of which inclined at 
$\alpha=30^o$ angle to the rotation axis of the star. See text for details.}
  \label{fig:c4-a30i10i70}
 \end{center}
\end{figure}

  We plot in figure~\ref{fig:c4-a30i10i70} results of our calculations
for the model with $\alpha=30^o,$ $\dot M_{ac}\simeq 4\cdot 10^{-8}$ 
M$_\odot$/yr, $i=10^o$ (left panel) and $i=70^o$ (right panel). Vertical
row of profiles in each panel corresponds to the following set of 
rotation period phases (from top to bottom): $\psi=0,$ 0.25, 0.5, 0.75.
Profiles were normalized to maximal intensity of the line at $\psi=0$
phase. It was assumed that accretion disk does not prevent to observe
the part of the star that situated below disk's midplane -- this is the
reason why some profiles have extended blue wing.

  Matter falls to the star with dipole magnetic field at the angle 
$\theta <90^o$ relative to its surface. In the absence of magnetic field 
oblique shock should arise in such situation, what means that: 1) the 
shock front is parallel to stellar surface; 2) velocity component $V_r$ 
which parallel to stellar radius is the pre-shock velocity $V_0.$ But 
bear in mind that accreted gas moves along magnetic field lines it also 
seems resonable to suppose that shock front is perpendicular to the 
magnetic field lines and therefore $V_0=V.$ To avoid discussion of this
problem we calculated profiles for both cases: solid lines in 
figure~\ref{fig:c4-a30i10i70} corresponds to profiles calculated for 
$V_0=V_r$ case and dashed -- for $V_0=V$ case.

  Such approach looks resonable at the moment because both types of
theoretical profiles differs from profiles of C\,IV\,1548 line in
spectra of CTTS presented in figures~\ref{fig:c4-obsprof-1},
\ref{fig:c4-obsprof-2}. Observed profiles have only one peak, maximum
of which is almost at zero velocity position. The only exception is
DR Tau: profile its C\,IV\,1548 line consists of two redshifted 
components but intensity of "high-velocity" component is larger 
than "low-velocity" one. Theoretical profiles calculated for
models with another values of $\alpha,$ $\dot M_{ac}$ and $i$ parameters 
have qulitatively the same shape as in figure~\ref{fig:c4-a30i10i70},
i.e. also can not reproduce observations.

  We suppose that the reason of the descripancy is too small divergency
of accreted gas stream lines within accretion zone that itself occupies only
$\sim 5\,\%$ of stellar surface (Romanova et al., 2003). If divergency of
the stream (and therefore magnetic field) lines within accretion zone would
be larger then it seems possible to obtain single-peak profile with extended
red wing similar to observed ones. Anyway our results indicate that magnetic
field of CTTSs is signifificantly non-dipole near stellar surface in
agreement with direct magnetic field measurements (Johns-Krull, 2007).

\section{Conclusion}

  We demonstrated that observed intensity and profiles of C\,IV
1550 doublet lines significantly differ from theoretical predictions 
based on the assumption that magnetic field of CTTSs near stellar surface
is close to dipole. We conclude therefore that geometry of CTTS's 
magnetic field near stellar surface is strongly non-dipole. Multipole
components of global magnetic field of young star or/and small-scale 
magnetic fields of active regions probably produce
large divergency of accreted gas stream lines within accretion zone
that presumbly can explain disagreement between the theory and observations.  

\medskip

{\it
We thank the LOC of the Symposium for the invitation, financial support
and hospitality.}

\end{document}